\documentclass[fleqn,usenatbib]{mn2e}

\usepackage{times}
\usepackage{mathptmx}

\usepackage[T1]{fontenc}
\usepackage{ae,aecompl}

\usepackage{graphicx}	
\usepackage{amsmath}	
\usepackage{amssymb}	

\title[Ni photoionization cross-sections]{Hot DA white dwarf model atmosphere calculations: Including improved Ni PI cross sections}

\author[S. P. Preval et al.]{S. P. Preval$^{1,2}$\thanks{E-mail:
simon.preval@strath.ac.uk}, M. A. Barstow$^{1}$, N. R. Badnell$^{2}$, I. Hubeny$^{3}$, J. B. Holberg$^{4}$ \\
$^{1}$Department of Physics and Astronomy, University of Leicester, University Road, Leicester, LE1 7RH \\
$^{2}$Department of Physics, University of Strathclyde, Glasgow, G4 0NG \\
$^{3}$Steward Observatory, University of Arizona, Tucson, AZ 85721, USA\\
$^{4}$Lunar and Planetary Laboratory, Sonett Space Sciences Building, University of Arizona, Tucson, AZ 85721, USA\\}

\date{Accepted XXX. Received YYY; in original form ZZZ}

\pubyear{2016}

\begin{document}
\label{firstpage}
\pagerange{\pageref{firstpage}--\pageref{lastpage}}
\maketitle

\begin{abstract}
To calculate realistic models of objects with Ni in their atmospheres, accurate atomic data for the relevant 
ionization stages needs to be included in model atmosphere calculations. In the context of white dwarf stars, we 
investigate the effect of changing the Ni {\sc iv}-{\sc vi} bound-bound and bound-free atomic data has on model 
atmosphere calculations. Models including PICS calculated with {\sc autostructure} show significant flux attenuation 
of up to $\sim 80$\% shortward of 180\AA\, in the EUV region compared to a model using hydrogenic PICS. 
Comparatively, models including a larger set of Ni transitions left the EUV, UV, and optical continua unaffected. We 
use models calculated with permutations of this atomic data to test for potential changes to measured metal abundances 
of the hot DA white dwarf G191-B2B. Models including {\sc autostructure} PICS were found to change the abundances of 
N and O by as much as $\sim 22$\% compared to models using hydrogenic PICS, but heavier species were relatively 
unaffected. Models including {\sc autostructure} PICS caused the abundances of N/O {\sc iv} and {\sc v} to diverge. 
This is because the increased opacity in the {\sc autostructure} PICS model causes these charge states to form 
higher in the atmosphere, moreso for N/O {\sc v}. Models using an extended line list caused significant changes 
to the Ni {\sc iv}-{\sc v} abundances. While both PICS and an extended line list cause changes in both synthetic 
spectra and measured abundances, the biggest changes are caused by using {\sc autostructure} PICS for Ni.
\end{abstract}




\section{Introduction}
The presence of metals in white dwarf (WD) atmospheres can have dramatic effects on both the structure of the atmosphere, and 
the observed effective temperature ($T_{\mathrm{eff}}$). For example, these effects have been demonstrated convincingly by 
\cite{barstow98a}. The authors determined the $T_{\mathrm{eff}}$ and surface gravity (log $g$) of several hot DA white dwarfs 
using a set of model atmosphere grids, which were either pure H \& He, or heavy metal-polluted. It was found that the 
$T_{\mathrm{eff}}$ determined using the pure H model grid was $\approx{4000-7000}$K higher than if a heavy metal-polluted 
model grid were used. Conversely, there was little to no difference in the measured log $g$ when using either model grid.

It can be inferred that the completeness of the atomic data supplied in model calculations might have a significant effect on 
the measured $T_{\mathrm{eff}}$. A study by \cite{chayer95a} considered the effects of radiative levitation on the observed 
atmospheric metal abundances at different $T_{\mathrm{eff}}$ and log $g$. In addition, these calculations were done using Fe 
data sets of varying line content. It was found that the number of transitions included in the calculation greatly affected 
the expected Fe abundance in the atmosphere (cf. \citealt{chayer95a}, Figure 11). This result implies that the macroscopic 
quantities determined in a white dwarf, such as metal abundance, are extremely sensitive to the input physics used to 
calculate the model grids. Therefore, this means that any atomic data that is supplied to the calculation must be as complete 
and accurate as possible in order to calculate the most representative model. While the \cite{chayer95a} study considered 
only the variations in observed Fe abundance, it is reasonable to assume that the set of atomic data supplied may also have 
an impact on the $T_{\mathrm{eff}}$ and log $g$ measured. 

Studies of white dwarf metal abundances, such as those of \cite{barstow98a,vennes01a,preval13a}, used model atmospheres 
incorporating the atomic data of \cite{kurucz92a} (Ku92 hereafter) in conjunction with photoionization cross-section (PICS) 
data from the Opacity Project (OP) for Fe, and approximate hydrogenic PICS for Ni. A more comprehensive dataset has since 
been calculated by \cite{kurucz11a} (Ku11 hereafter), containing a factor $\sim{10}$ more transitions and energy levels for 
Fe and Ni {\sc iv-vi} than its predecessor. In Table \ref{table:lines} we give a comparison of the number of Fe and Ni 
{\sc iv-vii} transitions available between Ku92 and Ku11. In both the Ku92 and Ku11 datasets, the energy levels etc were 
calculated using the Cowan Code \citep{cowanbook1981}. Based on the work discussed above, it is prudent to explore the 
differences between model atmospheres calculated using the Ku92 and Ku11 transition data, and the effect this has on 
measurements made using such models. 

\begin{table}
\renewcommand{\arraystretch}{1.1}
\centering
\caption{The number of lines present in the Ku92 and Ku11 atomic databases for Fe and Ni {\sc iv}-{\sc vii}.}
\begin{tabular}[H]{@{}lrr}
\hline
Ion & No of lines 1992 & No of lines 2011 \\
\hline
Fe {\sc iv}  &  1,776,984 &  14,617,228 \\
Fe {\sc v}   &  1,008,385 &   7,785,320 \\
Fe {\sc vi}  &    475,750 &   9,072,714 \\
Fe {\sc vii} &     90,250 &   2,916,992 \\
Ni {\sc iv}  &  1,918,070 &  15,152,636 \\
Ni {\sc v}   &  1,971,819 &  15,622,452 \\
Ni {\sc vi}  &  2,211,919 &  17,971,672 \\
Ni {\sc vii} &    967,466 &  28,328,012 \\
Total        & 10,420,643 & 111,467,026 \\
\hline
\end{tabular}
\label{table:lines}
\renewcommand{\arraystretch}{1.0}
\end{table}

Ni {\sc v} absorption features were first discovered in the hot DA white dwarfs G191-B2B and REJ2214-492 by \cite{holberg94a}
using high dispersion UV spectra from the International Ultraviolet Explorer (IUE). The authors derived Ni abundances of 
$\sim 1\times{10}^{-6}$ and $\sim 3\times{10}^{-6}$ as a fraction of H for G191-B2B and REJ2214-492 respectively. Compared to their 
measurements for Fe of $\sim 3\times{10}^{-5}$ and $\sim 1\times{10}^{-4}$ for G191-B2B and REJ2214-492 respectively, Ni is $\sim 3$\% of 
the Fe abundance for these stars. \cite{werner94a} also found Ni in two other hot DA white dwarfs, namely Feige 24, and 
RE0623-377, measuring Ni abundances of $1-5\times{10}^{-6}$ and $1-5\times{10}^{-5}$, respectively. More recently, 
\cite{preval13a} and \cite{rauch13a} have measured Ni abundances for G191-B2B. Using Ni {\sc iv} and {\sc v}, 
\cite{preval13a} found abundances of $3.24\times{10}^{-7}$ and $1.01\times{10}^{-6}$ respectively, while \cite{rauch13a} 
measured a single Ni abundance of $6\times{10}^{-7}$.

To date, there has been no attempt to include representative PICS for Ni in white dwarf model atmosphere 
calculations. It is for this reason that we choose to focus on Ni, and examine the effects of both including new PI cross 
sections, and transition data. We structure our paper as follows; We first describe the atomic data calculations required to 
generate PICS for Ni {\sc iv-vi}. Next, we describe the model atmosphere calculations performed, 
and the tests we conducted. We then discuss the results. Finally, we state our conclusions. 

\section{Atomic data calculations}
The PICS data presented in this paper were calculated using the distorted wave atomic collision package 
{\sc autostructure} \citep{badnell86a,badnell97a,badnell11a}. {\sc autostructure} is an atomic structure package that can 
model several aspects of an arbitrary atom/ion a priori, including energy levels, transition oscillator strengths, 
electron-impact excitation cross sections, PICS, and many others. {\sc autostructure} is supplied with a set of 
configurations describing the number of electrons and the quantum numbers occupied for a given atom or ion. The wavefunction 
$P_{nl}$ for a particular configuration is obtained by solving the one particle Schrodinger equation
\begin{equation}
\left[\frac{d^2}{dr^2}-\frac{l(l+1)}{r^2}+2V_{\mathrm{Eff}}(r)+E_{nl}\right]P_{nl}=0,
\end{equation}
where $n$ and $l$ are the principle and orbital angular momentum quantum numbers respectively, and $V_{\mathrm{Eff}}$ is an 
effective potential as described by \cite{eissner69a}, which accounts for the presence of other electrons. Scaling parameters
$\lambda_{nl}$ can be included to scale the radial coordinate, which are related to the effective charge `seen' by a 
particular valence electron, and typically has a value close to unity. The $\lambda_{nl}$ can be varied according to the 
task. For an example, the parameters may be varied to minimise an energy functional, or they may be varied such that the 
difference between the calculated energy levels and a set of observed energy levels is minimised. Three coupling schemes are 
available in calculating the wavefunctions, dependent upon the resolution required, and the type of problem being considered.
These are LS coupling (term resolved), intermediate coupling (IC, level resolved), or configuration average (CA, 
configuration resolved).

We use IC with the aim of reproducing the energy levels from Ku11 as closely as possible. The energy level structure is first
determined by running {\sc autostructure} with the configurations used by Ku11, listed in Table \ref{table:asconfig}. With 
this information, 17 $\lambda_{nl}$ are then specified for orbitals 1s to 6s set initially to unity. The 1s parameter is 
fixed as it does not converge in the presence of relativistic corrections. The parameters were then varied to give as close 
agreement to the Ku11 data as possible. The result of parameter variation is given in Table \ref{table:asparm}. 

\begin{table*}
\renewcommand{\arraystretch}{1.1}
\centering
\caption{Configurations prior to photoionization used in the {\sc autostructure} calculations. Configurations in bold typeset 
represent the ground state configuration. The columns $N_\mathrm{config}$, Levels, and Lines give the number of 
configurations used, and the number of energy levels and transitions generated respectively.}
\begin{tabular}[H]{@{}lllll}
\hline
Ion & Configurations & $N_\mathrm{config}$ & Levels & Lines \\
\hline
Ni {\sc iv} & $\mathbf{3d^7}$, $3d^6 4s$, $3d^6 5s$,$3d^6 6s$, $3d^6 7s$, $3d^6 8s$ &  85 & 37,860 & 32,416,571 \\
            & $3d^6 9s      $, $3d^5 4s^2   $, $3d^5 4s5s  $, $3d^5 4s6s  $, $3d^5 4s7s  $, $3d^5 4s8s   $ & & & \\
            & $3d^5 4s 9s   $, $3d^4 4s^2 5s$, $3d^6 4d     $, $3d^6 5d    $, $3d^6 6d    $, $3d^6 7d     $ & & & \\
            & $3d^6 8d      $, $3d^6 9d    $, $3d^5 4s4d  $, $3d^5 4s5d  $, $3d^5 4s6d  $, $3d^5 4s7d   $ & & & \\
            & $3d^5 4s 8d   $, $3d^5 4s 9d  $, $3d^4 4s^2 4d$, $3d^5 4p^2   $, $3d^6 5g    $, $3d^6 6g     $ & & & \\
            & $3d^6 7g      $, $3d^6 8g    $, $3d^6 9g    $, $3d^5 4s 5g  $, $3d^5 4s 6g  $, $3d^5 4s 7g   $ & & & \\
            & $3d^5 4s 8g   $, $3d^5 4s 9g  $, $3d^6 7i    $, $3d^6 8i    $, $3d^6 9i    $, $3d^5 4s 7i   $ & & & \\
            & $3d^5 4s 8i   $, $3d^5 4s 9i$ & & & \\
            & $3d^6 4p      $, $3d^6 5p    $, $3d^6 6p    $, $3d^6 7p    $, $3d^6 8p    $, $3d^6 9p     $ & & & \\
            & $3d^5 4s 4p   $, $3d^5 4s 5p  $, $3d^5 4s 6p  $, $3d^5 4s 7p  $, $3d^5 4s 8p  $, $3d^5 4s 9p   $ & & & \\
            & $3d^4 4s^2 4p $, $3d^6 4f    $, $3d^6 5f    $, $3d^6 6f    $, $3d^6 7f    $, $3d^6 8f     $ & & & \\
            & $3d^6 9f      $, $3d^5 4s 4f  $, $3d^5 4s 5f  $, $3d^5 4s 6f  $, $3d^5 4s 7f  $, $3d^5 4s 8f   $ & & & \\
            & $3d^5 4s 9f   $, $3d^4 4s^2 4f$, $3d^6 6h    $, $3d^6 7h    $, $3d^6 8h    $, $3d^6 9h     $ & & & \\
            & $3d^5 4s 6h   $, $3d^5 4s 7h  $, $3d^5 4s 8h  $, $3d^5 4s 9h  $, $3d^6 8k    $, $3d^6 9k     $ & & & \\
            & $3d^5 4s 8k   $, $3d^5 4s 9k  $, $3p^5 3d^8$ & & & \\
Ni {\sc v}  & $\mathbf{3d^6}$, $3d^5 4d$,  $3d^5 5d$,      $3d^5 6d$, $3d^5 7d $, $3d^5 8d $ &  87 & 37,446 & 34,066,259 \\
            & $3d^5 9d      $, $3d^5 10d   $, $3d^4 4s4d   $, $3d^4 4s5d  $, $3d^4 4s6d  $, $3d^4 4s7d   $ & & & \\
            & $3d^4 4s 8d   $, $3d^4 4s9d  $, $3d^4 4s10d  $, $3d^5 4s    $, $3d^5 5s    $, $3d^5 6s     $ & & & \\
            & $3d^5 7s      $, $3d^5 8s    $, $3d^5 9s     $, $3d^5 10s   $, $3d^4 4s^2  $, $3d^4 4s5s   $ & & & \\
            & $3d^4 4s 6s   $, $3d^4 4s7s  $, $3d^4 4s8s   $, $3d^4 4s9s  $, $3d^4 4s10s $, $3d^5 5g     $ & & & \\
            & $3d^5 6g      $, $3d^5 7g    $, $3d^5 8g     $, $3d^5 9g    $, $3d^4 4s5g  $, $3d^4 4s6g   $ & & & \\
            & $3d^4 4s 7g   $, $3d^4 4s8g  $, $3d^4 4s9g   $, $3d^5 7i    $, $3d^5 8i    $, $3d^5 9i     $ & & & \\
            & $3d^4 4s 7i   $, $3d^4 4s8i  $, $3d^4 4s9i   $, $3d^4 4p^2$                                  & & & \\
            & $3d^5 4p      $, $3d^5 5p    $, $3d^5 6p     $, $3d^5 7p    $, $3d^5 8p    $, $3d^5 9p     $ & & & \\
            & $3d^5 10p     $, $3d^4 4s4p  $, $3d^4 4s5p   $, $3d^4 4s6p  $, $3d^4 4s7p  $, $3d^4 4s8p   $ & & & \\
            & $3d^4 4s 9p   $, $3d^4 4s10p $, $3d^3 4s^2 4p$, $3d^5 4f   $, $3d^5 5f    $, $3d^5 6f     $ & & & \\
            & $3d^5 7f      $, $3d^5 8f    $, $3d^5 9f    $, $3d^5 10f   $, $3d^4 4s4f  $, $3d^4 4s5f   $ & & & \\
            & $3d^4 4s 6f   $, $3d^4 4s7f  $, $3d^4 4s8f  $, $3d^4 4s9f  $, $3d^5 6h    $, $3d^5 7h     $ & & & \\
            & $3d^5 8h      $, $3d^5 9h    $, $3d^4 4s6h  $, $3d^4 4s7h  $, $3d^4 4s8h  $, $3d^4 4s9h   $ & & & \\
            & $3d^5 8k      $, $3d^5 9k    $, $3d^4 4s8k  $, $3d^4 4s9k  $, $3p^5 3d^7$ & & & \\
Ni {\sc vi} & $\mathbf{3d^5}$, $3d^4 4d$, $3d^4 5d$, $3d^4 6d$, $3d^4 7d$, $3d^4 8d$ & 122 & 29,366 & 42,412,822 \\
            & $3d^4 9d     $, $3d^4 10d   $, $3d^3 4s 4d  $, $3d^3 4s5d  $, $3d^3 4s6d  $, $3d^3 4s7d   $ & & & \\
            & $3d^3 4s 8d  $, $3d^3 4s 9d  $, $3d^3 4s 10d $, $3d^2 4s^2 4d  $, $3d^2 4s^2 5d  $, $3d^2 4s^2 6d   $ & & & \\
            & $3d^2 4s^2 7d  $, $3d^2 4s^2 8d   $, $3d^2 4s^2 9d  $, $3d^2 4s^2 10d $, $3d^4 4s    $, $3d^4 5s     $ & & & \\
            & $3d^4 6s     $, $3d^4 7s    $, $3d^4 8s    $, $3d^4 9s    $, $3d^4 10s   $, $3d^3 4s^2    $ & & & \\
            & $3d^3 4s 5s  $, $3d^3 4s 6s  $, $3d^3 4s 7s  $, $3d^3 4s8s  $, $3d^3 4s9s  $, $3d^3 4s10s  $ & & & \\
            & $3d^2 4s^2 5s   $, $3d^2 4s^2 6s  $, $3d^2 4s^2 7s  $, $3d^2 4s^2 8s$, $3d^2 4s^2 9s$, $3d^2 4s^2 10s$ & & & \\
            & $3d^4 5g     $, $3d^4 6g    $, $3d^4 7g    $, $3d^4 8g    $, $3d^4 9g    $, $3d^4 10g    $ & & & \\
            & $3d^3 4s 5g  $, $3d^3 4s 6g  $, $3d^3 4s 7g  $, $3d^3 4s8g  $, $3d^3 4s9g  $, $3d^3 4s10g  $ & & & \\
            & $3d^4 7i     $, $3d^4 8i    $, $3d^4 9i    $, $3d^3 4s7i  $, $3d^3 4s8i  $, $3d^3 4s9i $, $3d^3 4p^2$ & & & \\
            & $3d^4 4p     $, $3d^4 5p    $, $3d^4 6p    $, $3d^4 7p    $, $3d^4 8p    $, $3d^4 9p     $ & & & \\
            & $3d^4 10p    $, $3d^4 11p   $, $3d^3 4s 4p  $, $3d^3 4s5p  $, $3d^3 4s6p  $, $3d^3 4s7p   $ & & & \\
            & $3d^3 4s 8p  $, $3d^3 4s 9p  $, $3d^3 4s 10p $, $3d^3 4s11p $, $3d^2 4s^2 4p  $, $3d^2 4s^2 5p   $ & & & \\
            & $3d^2 4s^2 6p$, $3d^2 4s^2 7p$, $3d^2 4s^2 8p  $, $3d^2 4s^2 9p  $, $3d^2 4s^2 10p $, $3d^2 4s^2 11p$ & & & \\
            & $3d^4 4f     $, $3d^4 5f    $, $3d^4 6f    $, $3d^4 7f    $, $3d^4 8f    $, $3d^4 9f     $ & & & \\
            & $3d^4 10f    $, $3d^4 11f   $, $3d^3 4s 4f  $, $3d^3 4s5f  $, $3d^3 4s6f  $, $3d^3 4s7f   $ & & & \\
            & $3d^3 4s 8f  $, $3d^3 4s 9f  $, $3d^3 4s 10f $, $3d^3 4s11f $, $3d^2 4s^2 4f  $, $3d^2 4s^2 5f   $ & & & \\
            & $3d^2 4s^2 6f  $, $3d^2 4s^2 7f  $, $3d^2 4s^2 8f$, $3d^2 4s^2 9f$, $3d^2 4s^2 10f $, $3d^2 4s^2 11f$ & & & \\
            & $3d^4 6h     $, $3d^4 7h    $, $3d^4 8h    $, $3d^4 9h    $, $3d^3 4s6h  $, $3d^3 4s7h   $ & & & \\
            & $3d^3 4s 8h  $, $3d^3 4s 9h  $, $3d^4 8k    $, $3d^4 9k    $, $3d^3 4s8k  $, $3d^3 4s9k$, $3p^5 3d^6$ & & & \\
\hline
\end{tabular}
\label{table:asconfig}
\renewcommand{\arraystretch}{1.0}
\end{table*}

\begin{table}
\renewcommand{\arraystretch}{1.1}
\centering
\caption{Calculated IC scaling parameters from {\sc autostructure}.}
\begin{tabular}[H]{@{}llll}
\hline
Orbital & Ni {\sc iv} & Ni {\sc v} & Ni {\sc vi} \\
\hline
$2s$ & 1.31391 & 1.31355 & 1.31487 \\
$2p$ & 1.12294 & 1.12144 & 1.11996 \\
$3s$ & 1.09887 & 1.11219 & 1.12750 \\
$3p$ & 1.05876 & 1.07206 & 1.08779 \\
$3d$ & 1.06828 & 1.09148 & 1.10379 \\
$4s$ & 1.13492 & 1.15167 & 1.19770 \\
$4p$ & 0.91520 & 0.90669 & 0.92207 \\
$4d$ & 1.40156 & 1.48359 & 1.49036 \\
$4f$ & 1.08824 & 1.03771 & 1.10538 \\
$5s$ & 1.03946 & 1.03552 & 1.25294 \\
$5p$ & 0.99302 & 0.96743 & 1.00439 \\
$5d$ & 1.09409 & 1.08876 & 1.15671 \\
$5f$ & 1.06373 & 1.01089 & 1.01693 \\
$5g$ & 1.37199 & 1.17219 & 1.83712 \\
$6s$ & 1.02002 & 1.01140 & 1.05828 \\
\hline
\end{tabular}
\label{table:asparm}
\renewcommand{\arraystretch}{1.0}
\end{table}

By comparing Table \ref{table:lines} with Table \ref{table:asconfig}, it can be seen that the latter has more transitions 
than the former. This is because the transitions listed in Ku11 are limited by their strength. Any observed/well known 
transitions with log $gf<-9.99$, or predicted transitions with a log $gf<-7.5$ were omitted. In addition, the Ku11 database 
omitted radiative transitions between two autoionizing levels. We do the same. After calculating a set of scaling parameters 
for a particular ion, the accompanying PICS can be obtained. In order to see what potential effect (if any) replacing the 
hydrogenic PICS with more realistic data would have, we limited our calculations to considered direct 
photoionization (PI) only, neglecting resonances from photoexcitation/autoionization. The final photoionized configurations 
used in calculating the PICS were constructed by removing the outer most electron from each configuration in Table \ref{table:asconfigpi}. These PI
configurations are listed in Table \ref{table:asconfigpi}. The PICS are evaluated for a table of 50 logarithmically spaced 
ejected electron energies spanning 0 to 100 Ryd. The PICS in the ejected electron energy frame are then linearly 
interpolated to the incident photon energy frame using two point interpolation.

\begin{table}
\renewcommand{\arraystretch}{1.1}
\centering
\caption{Final photoionized configurations used in the {\sc autostructure} calculations.}
\begin{tabular}[H]{@{}ll}
\hline
Ion & Configurations \\
\hline
Ni {\sc iv} & $3p^5 3d^7$, $3d^6$, $3d^5 4s$, $3d^5 4p$, $3d^4 4s^2$  \\
Ni {\sc v}  & $3p^5 3d^6$, $3d^5$, $3d^4 4s$, $3d^4 4p$, $3d^3 4s^2$  \\
Ni {\sc vi} & $3d^4$, $3p^5 3d^5$, $3d^3 4s$, $3d^3 4p$, $3d^2 4s^2$  \\
\hline
\end{tabular}
\label{table:asconfigpi}
\renewcommand{\arraystretch}{1.0}
\end{table}

{\sc autostructure} employs the distorted wave method, which is an approximation. The calculations performed for the 
OP were done using an R-Matrix approach, which can potentially give the most accurate result in calculating PICS. Furthermore,
the R-Matrix method automatically includes resonances arising from photoexcitation/autoionization, and interference between these
two processes. In {\sc autostructure}, when the resonances are included separately, interference effects are neglected. The downside
to using an R-Matrix calculation is the length of time and computer resources required to perform the calculation. 

With regards to accuracy, \cite{seaton2004a} calculated term-resolved PI calculations for Fe {\sc viii} to Fe {\sc xiii} using {\sc autostructure}. The authors then
replaced data calculated for the OP, which consisted of R-Matrix plus {\sc superstructure} \citep{eissner69a} data for these ions, and re-evalulated the Rosseland Means
for a solar mixture. The Rosseland means calculated using the {\sc autostructure} data were found to be close to those calculated with the OP data.

This implies that the distorted wave method is a good indicator of the potential effects of including new data. Therefore, it is instructive to perform
such calculations before committing to a large R-Matrix calculation.

\section{Stellar atmosphere calculations}\label{sec:stelatcal}
All model atmospheres in this work were calculated using the non-local thermodynamic equilibrium (NLTE) stellar atmospheres 
code {\sc tlusty} \citep{hubeny88a,hubeny95a}, version 201. The models were then synthesised using {\sc synspec} 
\citep{hubeny11a}. {\sc tlusty} benefits from the hybrid CL/ALI method, which combines the Complete Linearisation (CL) and 
Accelerated Lambda Iteration (ALI) methods in order to accelerate the rate of convergence of a model. {\sc tlusty} has two 
methods in which to treat heavy metal opacity, namely opacity distribution functions (ODF), and opacity sampling (OS). 
Nominally, OS is a Monte-Carlo sampling method. However, in the limit of high resolution, OS is an exact method for 
accounting for opacity. We use OS, and specify a resolution of 5 fiducial\footnote{the Doppler width for Fe absorption 
features at $T_{\mathrm{eff}}$.} Doppler widths.

\begin{figure}
\begin{centering}
\includegraphics[width=80mm]{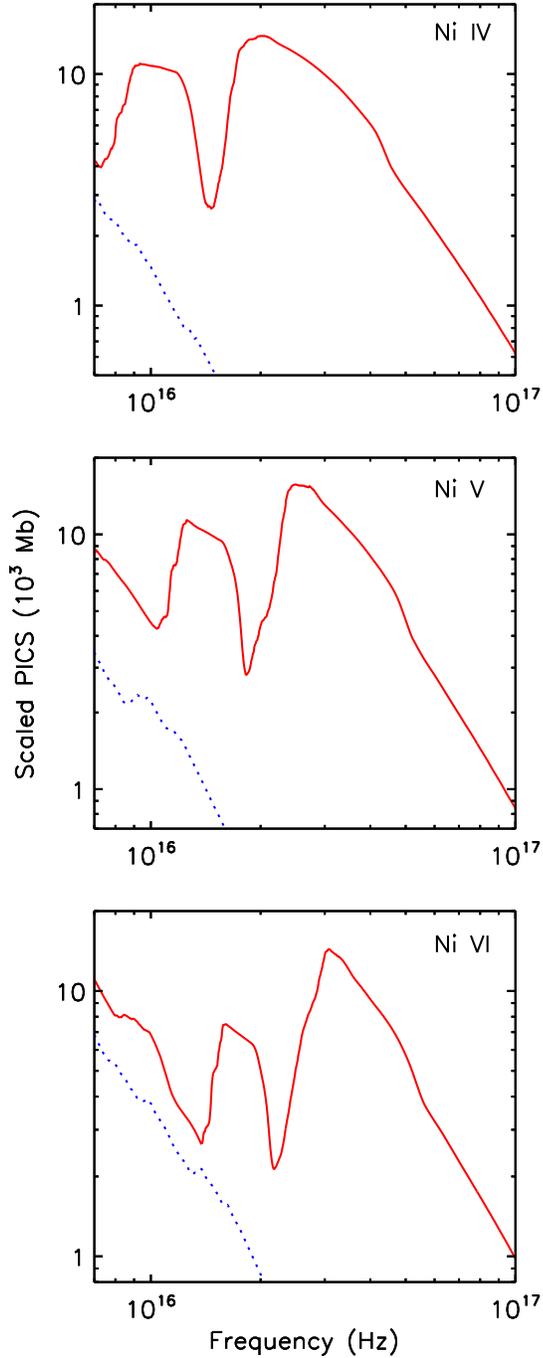}
\caption{Plot of summed supercross-sections for Ni {\sc iv-vi} calculated using autostructure (solid red) and a hydrogenic approximation (dotted blue). Prior to addition, each supercross-section was weight by a boltzmann
factor and the statistical weight of the superlevel concerned.}
\label{fig:xsects}
\end{centering}
\end{figure}

As the Ku11 data contains more energy levels than Ku92, we constructed new model ions for Ni {\sc iv-vi} according to 
the prescription of \cite{anderson89a}, however, superlevels were calculated using energies of either even or odd parity. If 
we created superlevels with a mixture of even or odd levels, then we would have to consider transitions between levels within 
the superlevel (cf. \citealt{hubeny95a}). The Ni {\sc iv-vi} ions have 73, 90, and 75 superlevels respectively. The PICS for 
each superlevel $\bar{\sigma}_{\mathrm{PI}}(E_{\gamma})$ as a function of photon energy $E_{\gamma}$ were calculated as an 
average of the PICS for each individual level $\sigma_{i}(E)$ used to create the superlevel, weighted by the statistical 
weights of each level $g_{i}$. This can be written as
\begin{equation}
\bar{\sigma}_{\mathrm{PI}}(E_{\gamma})=\frac{\sum_{i=1}^{N}{\sigma_{i}(E_{\gamma})g_{i}}}{\sum_{i=1}^{N}{g_{i}}},
\end{equation}
where $N$ is the number of levels used to form the superlevel. We refer to these summed PICS as supercross-sections 
hereafter. In Figure \ref{fig:xsects} we have plotted the total supercross-sections for Ni {\sc iv-vi} for both the hydrogenic and autostructure 
cases. Prior to summation, each supercross-section was multiplied by a boltzmann constant. It can be seen that the total 
supercross-sections calculated in {\sc autostructure} are far larger than their hydrogenic counterparts.

As there are two different Ni line lists (Ku92 and Ku11), and two different sets of PICS (hydrogenic and those 
calculated with {\sc autostructure}), there are four different combinations that we can test. Therefore, we calculated four 
different models in NLTE. We refer to these as Models 1, 2, 3, and 4. In Model 1, we use the Ku92 transitions 
and hydrogenic PICS. In Model 2, we use the Ku11 transitions and hydrogenic PICS. In Model 3, we use the 
Ku92 transitions and {\sc autostructure} PICS. Finally, in Model 4, we use the Ku11 transitions and 
{\sc autostructure} PICS. In all four cases, we based the models on a G191-B2B like atmosphere, and calculated the 
models with $T_{\mathrm{eff}}=52500$K, log $g=7.53$ \citep{barstow03b}, and metal abundances listed in Table 
\ref{table:atmosabun}. These abundances were taken from \cite{preval13a}, measured in their analysis of the hot DA white 
dwarf G191-B2B. In the case where there was more than one ionization stage considered, we used the abundances with the 
smallest uncertainty. Listed in Table \ref{table:opacions} are the model ions used in {\sc tlusty}, along with the number of 
superlevels included.

\begin{table}
\renewcommand{\arraystretch}{1.1}
\centering
\caption{Metal abundances used in calculating the four model atmospheres described in text as a fraction of H. These 
abundances originate from \protect\cite{preval13a}, where the values with the lowest statistical uncertainty were used.}
\begin{tabular}[H]{@{}cc}
\hline
Metal & Abundance X/H \\
\hline
He & $1.00\times{10}^{-5}$ \\
C  & $1.72\times{10}^{-7}$ \\
N  & $2.16\times{10}^{-7}$ \\
O  & $4.12\times{10}^{-7}$ \\
Al & $1.60\times{10}^{-7}$ \\
Si & $3.68\times{10}^{-7}$ \\
P  & $1.64\times{10}^{-8}$ \\
S  & $1.71\times{10}^{-7}$ \\
Fe & $1.83\times{10}^{-6}$ \\
Ni & $1.01\times{10}^{-6}$ \\
\hline
\end{tabular}
\label{table:atmosabun}
\renewcommand{\arraystretch}{1.0}
\end{table}

\begin{table}
\renewcommand{\arraystretch}{1.1}
\centering
\caption{List of model ions and the number of levels used in model atmosphere calculations described in the text. Ions marked 
with * were treated approximately as single level ions by {\sc tlusty}. For Ni, the number of levels outside and inside the 
brackets correspond to the number of levels for the Ku92 and Ku11 model ions, respectively.}
\begin{tabular}[H]{@{}lc}
\hline
Ion & Nsuperlevels \\
\hline
H  {\sc i}    & 9  \\ 
H  {\sc ii}*  & 1  \\
He {\sc i}    & 24 \\
He {\sc ii}   & 20 \\
He {\sc iii}* & 1  \\
C  {\sc iii}  & 23 \\
C  {\sc iv}   & 41 \\ 
C  {\sc v}*   & 1  \\
N  {\sc iii}  & 32 \\
N  {\sc iv}   & 23 \\ 
N  {\sc v}    & 16 \\
N  {\sc vi}*  & 1  \\
O  {\sc iv}   & 39 \\ 
O  {\sc v}    & 40 \\
O  {\sc vi}   & 20 \\ 
O  {\sc vii}* & 1  \\
Al {\sc iii}  & 23 \\
Al {\sc iv}*  & 1  \\
Si {\sc iii}  & 30 \\
Si {\sc iv}   & 23 \\ 
Si {\sc v}*   & 1  \\
P  {\sc iv}   & 14 \\
P  {\sc v}    & 17 \\ 
P  {\sc vi}*  & 1  \\
S  {\sc iv}   & 15 \\
S  {\sc v}    & 12 \\ 
S  {\sc vi}   & 16 \\
S  {\sc vii}* & 1  \\
Fe {\sc iv}   & 43 \\ 
Fe {\sc v}    & 42 \\
Fe {\sc vi}   & 32 \\ 
Fe {\sc vii}* & 1  \\
Ni {\sc iv}   & 38 (73) \\
Ni {\sc v}    & 48 (90) \\ 
Ni {\sc vi}   & 42 (75) \\
Ni {\sc vii}* & 1  \\
\hline
\end{tabular}
\label{table:opacions}
\renewcommand{\arraystretch}{1.0}
\end{table}

\subsection{Spectral Energy Distribution}
For this comparison, we considered the differences between the spectral energy distributions (SED) of each model. For each 
model, we synthesize three spectra covering the extreme ultraviolet (EUV), the ultraviolet (UV), and the optical regions. We 
then calculated the residual between models 1 and 2, 1 and 3, and 1 and 4 using the equation
\begin{equation}\label{eq:res}
\mathrm{Residual}=\frac{F_{\mathrm{i}}-F_{\mathrm{1}}}{F_{\mathrm{1}}},
\end{equation}
where $F_{\mathrm{1}}$ is the flux for model 1, and $F_{\mathrm{i}}$ is the flux for model 2, 3, or 4. 

\subsection{Abundance variations}
For this comparison, we wanted to examine the differences between abundances measured for G191-B2B when using each of the 
four models described above. Using a similar method to \cite{preval13a}, we measured the abundances for G191-B2B using all
four of the models described above. The observational data for G191-B2B consists of three high S/N spectra constructed by 
co-adding multiple datasets. The first spectrum uses data from the Far Ultraviolet Spectroscopic Explorer (FUSE) spanning 
910-1185\AA, and the other two use data from the Space Telescope Imaging Spectrometer (STIS) aboard the Hubble Space 
Telescope (HST) spanning 1160-1680\AA\, and 1625-3145\AA, respectively. A full list of the data sets used, and the coaddition 
procedure, is given in detail in \cite{preval13a}.

The model grids for each metal was constructed by using {\sc synspec}. {\sc synspec} takes a starting model converged assuming NLTE, and 
is able to calculate a spectrum for smaller or larger metal abundances by stepping away in LTE. We used the X-ray spectral 
package {\sc xspec} \citep{arnaud96a} to measure the abundances. {\sc xspec} takes a grid of models and observational data 
and interpolates between these models using a chi square ($\chi^2$) minimisation procedure.  {\sc xspec} is unable to use 
observational data with a large number of data points. To remedy this, we isolate individual absorption features for various 
ions and then use {\sc xspec} to measure the abundances. A full list of the absorption features used and the sections of 
spectrum extracted is given in Table 9 in \cite{preval13a}. In addition to this list, we also include measurements of the O 
{\sc v} abundance using the excited transition with wavelength 1371.296\AA.

\section{Results and discussion.}
Here we discuss the results obtained from the four models calculated using permutations of the Ku92 and Ku11 atomic data,
and the hydrogenic and {\sc autostructure} cross-section data. Models 1, 2, 3, and 4 used Ku92/hydrogenic, Ku11/hydrogenic,
Ku92/{\sc autostructure}, and Ku11/{\sc autostructure} respectively.

\subsection{SED variations}
In this subsection we discuss the differences between spectra synthesised for the four models described above.

\subsubsection{EUV}
\begin{figure*}
\begin{centering}
\includegraphics[width=140mm]{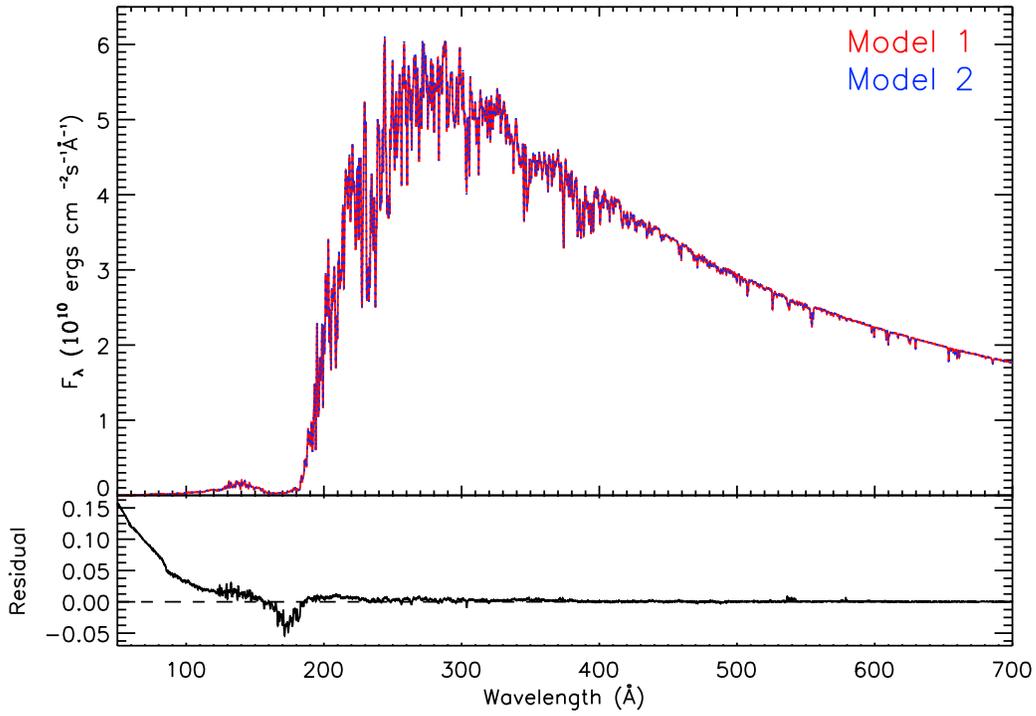}
\caption{Plot of the EUV region covering 50-700\AA\, synthesised for models 1 (red solid) and 2 (blue dotted). On the bottom is 
a plot of the residual between the two models. The dashed line indicates a residual of zero, or no difference.}
\label{fig:g191euvcomp1}
\end{centering}
\end{figure*}

\begin{figure*}
\begin{centering}
\includegraphics[width=140mm]{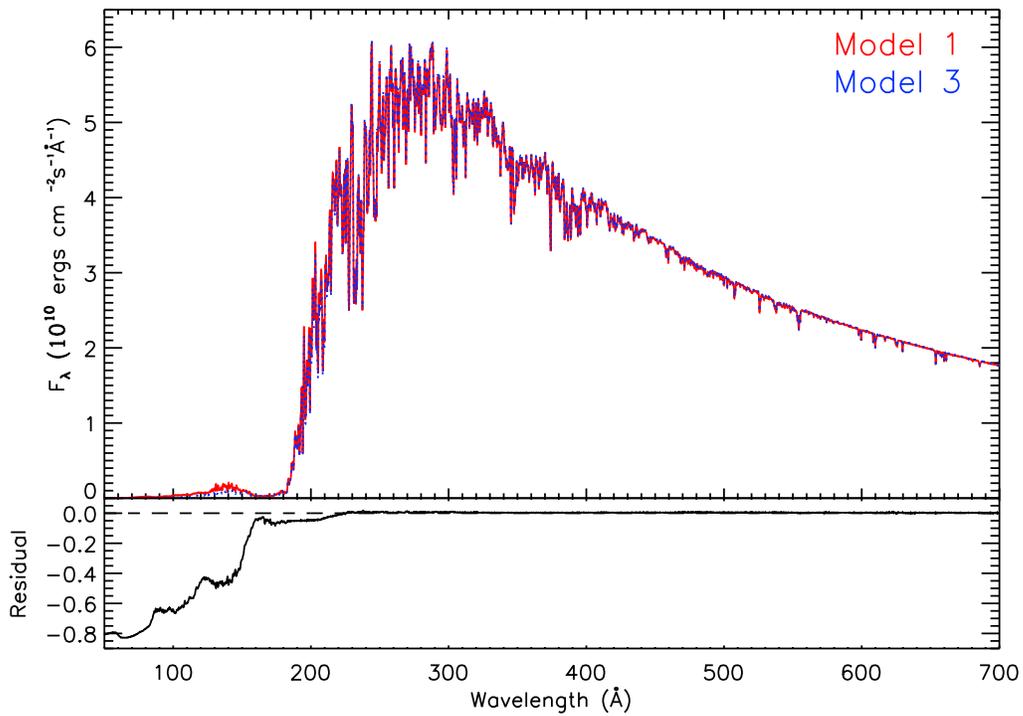}
\caption{Same as Figure \ref{fig:g191euvcomp1}, but for models 1 and 3.}
\label{fig:g191euvcomp2}
\end{centering}
\end{figure*}

Of all the spectral regions, the EUV undergoes the most dramatic changes. However, the EUV region appears to be relatively 
insensitive to whether Ku92 or Ku11 is used. In Figure \ref{fig:g191euvcomp1} we have plotted the EUV region for models 1 and
2, along with the residual between these two models as defined in text. Below 200\AA\, the flux of model 2 appears to increase 
as wavelength decreases, being $\sim 15$\% larger by 50\AA. This may be due to how the superlevels are partitioned in the model 
calculation rather than a decrease in opacity from the Ku11 line list. In Figure \ref{fig:g191euvcomp2} we plot the same 
region, but with models 1 and 3. Significant changes occur below 180\AA, with the flux of model 3 being greatly attenuated, 
reaching a maximum of $\sim 80$\% with respect to model 1. This is indicative of a larger opacity due to the {\sc autostructure} 
PICS for Ni. Model 4 showed a combination of effects from models 2 and 3.

\subsubsection{UV}
\begin{figure*}
\begin{centering}
\includegraphics[width=140mm]{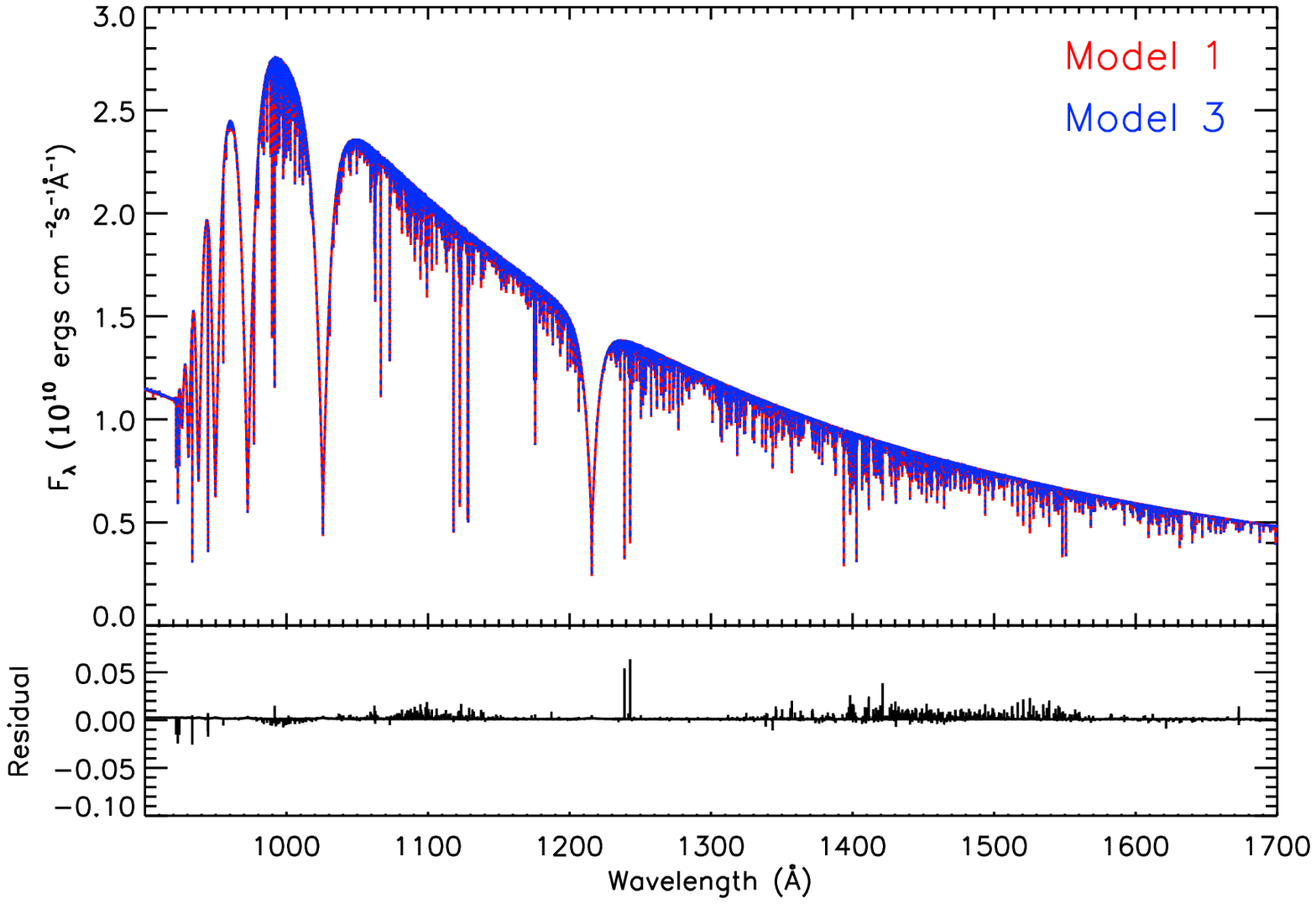}
\caption{Plot of the UV region covering 910-1700\AA\, synthesised for models 1 (red solid) and 2 (blue dotted). On the bottom 
is a plot of the residual between the two models. The dashed line indicates a residual of zero, or no difference.}
\label{fig:g191uvcomp1}
\end{centering}
\end{figure*}

\begin{figure*}
\begin{centering}
\includegraphics[width=140mm]{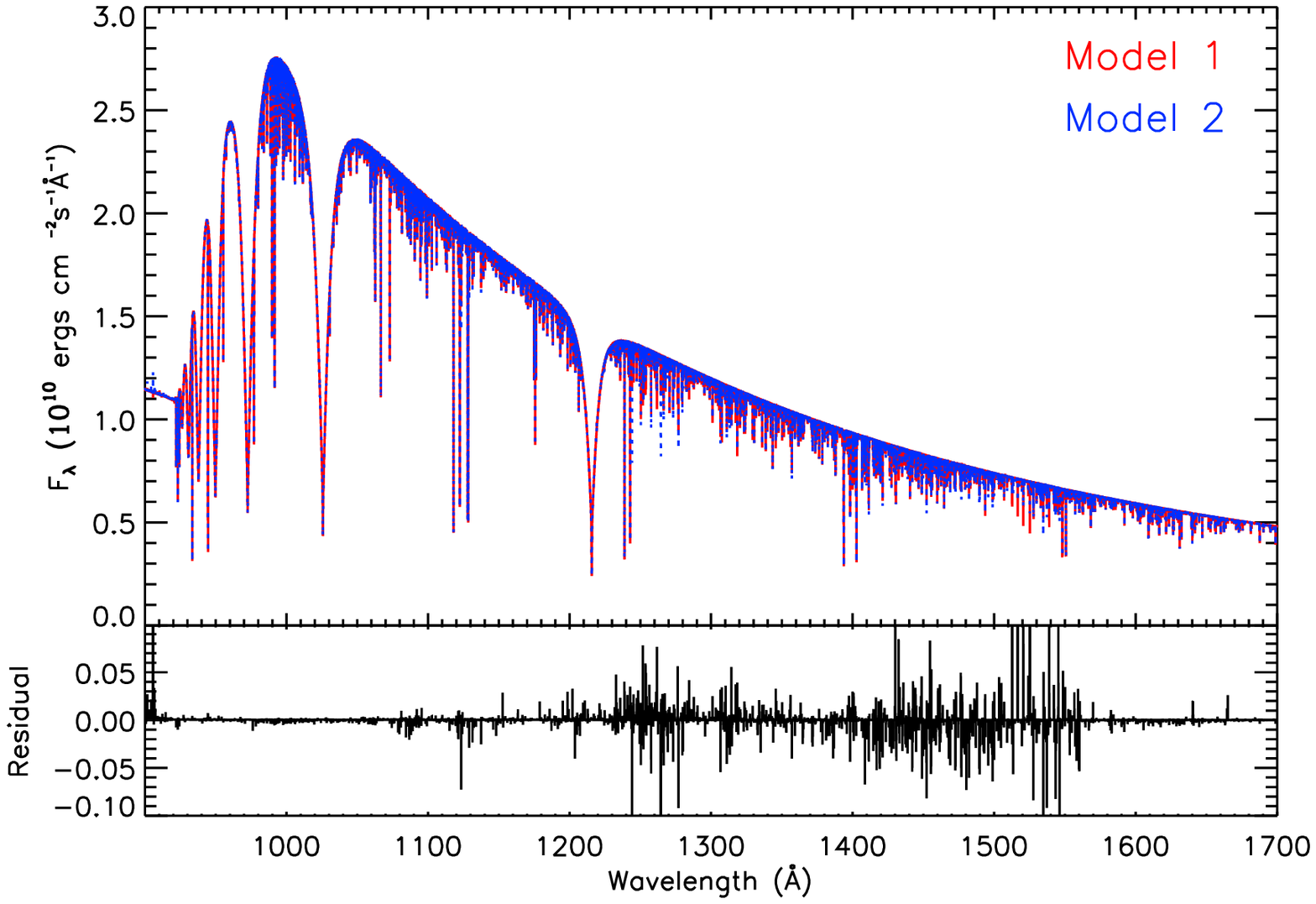}
\caption{Same as Figure \ref{fig:g191uvcomp1}, but for models 1 (red solid) and 3 (blue dotted).}
\label{fig:g191uvcomp2}
\end{centering}
\end{figure*}

In the case of the UV region, not a lot changes when using the {\sc autostructure} PICS. In Figure 
\ref{fig:g191uvcomp1} we have plotted synthetic spectra for Models 1 and 3 in the UV region. It can be seen that changes are 
limited to absorption features only, with the vast majority only changing depth by $\sim 3$\%. The obvious exception to this is 
the N {\sc v} doublet near 1240\AA, where the depth has changed by $\sim 5-6$\%. In Figure \ref{fig:g191uvcomp2} we have now 
plotted models 1 and 2. Again, changes are limited to absorption features, but these are now far more pronounced, with depth 
changes of up to and beyond $10$\%. These features can be attributed to Ni, and a few lighter metals, the abundances of which 
we discuss later. Again, Model 4 showed a combination of the changes seen in models 2 and 3.

\subsubsection{Optical}
\begin{figure*}
\begin{centering}
\includegraphics[width=140mm]{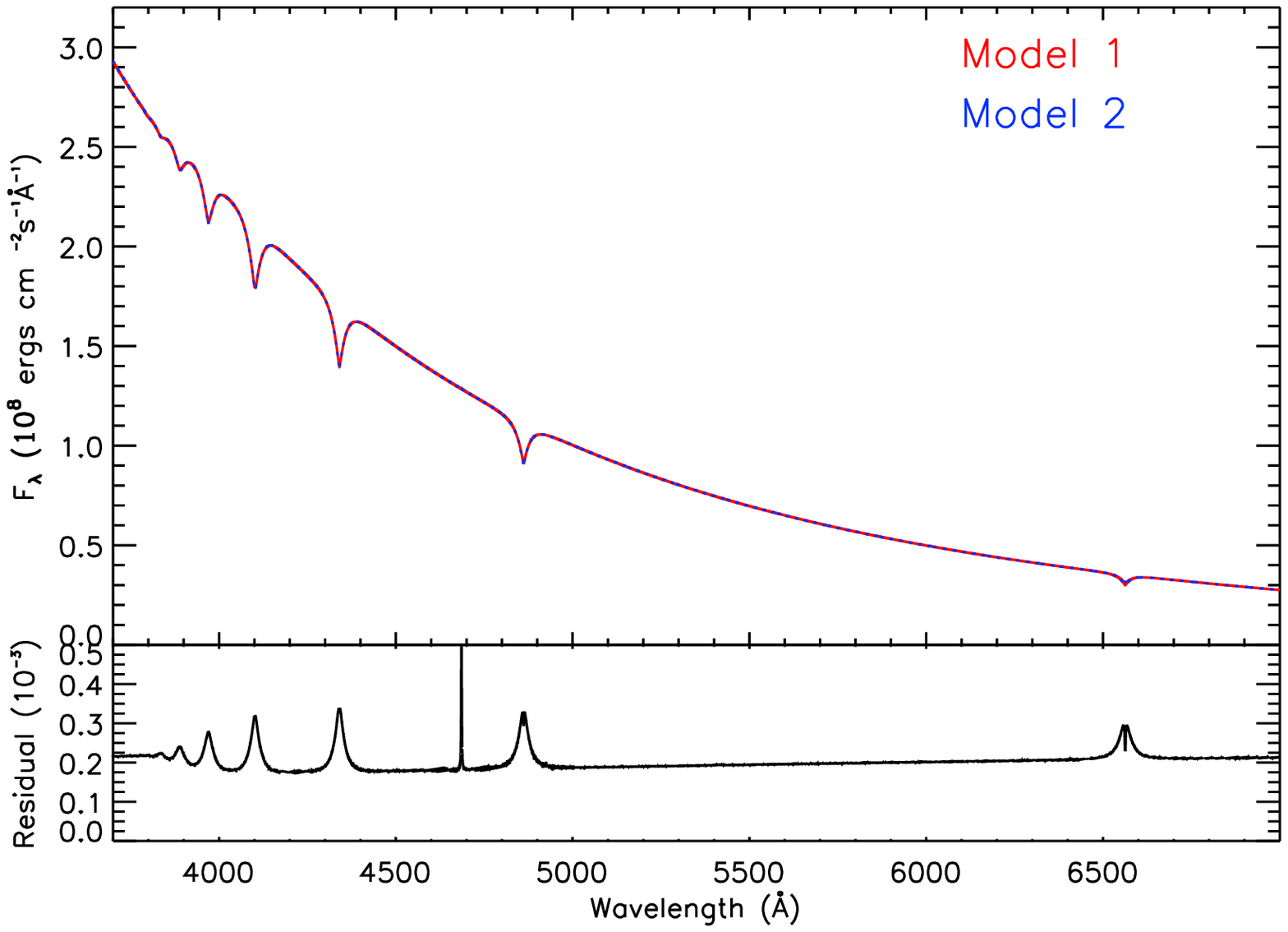}
\caption{Plot of the optical region covering 3800-7000\AA\, synthesised for models 1 (red solid) and 2 (blue dotted). On the 
bottom is a plot of the residual between the two models. The sharp residual at ~4690\AA\, is due to the He II 4860.677\AA\, 
transition.}
\label{fig:g191optcomp1}
\end{centering}
\end{figure*}

\begin{figure*}
\begin{centering}
\includegraphics[width=140mm]{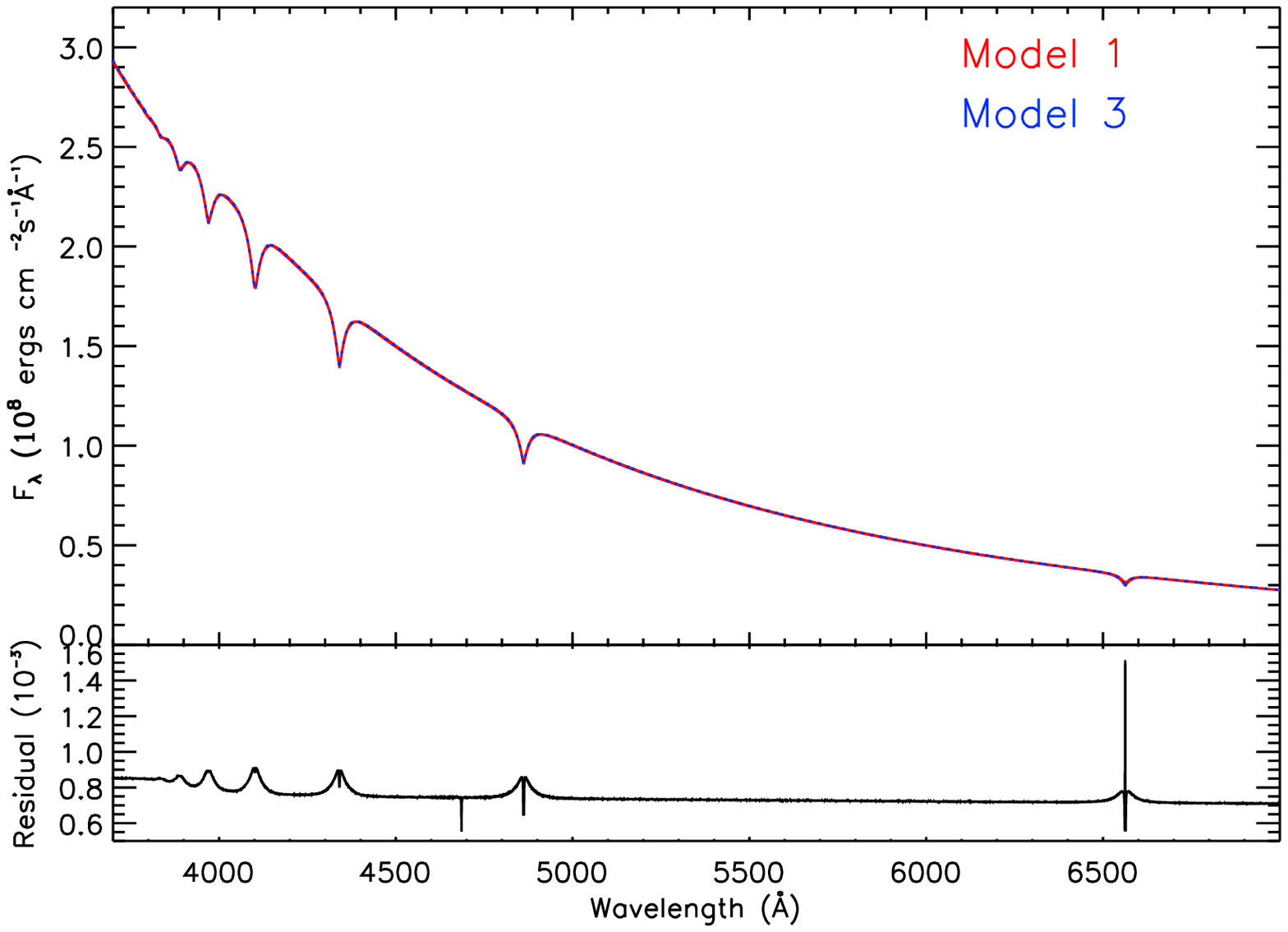}
\caption{Same as Figure \ref{fig:g191optcomp1}, but for models 1 (red solid) and 3 (blue dotted).}
\label{fig:g191optcomp2}
\end{centering}
\end{figure*}

Very little to no change occurs in the optical region, regardless of line list or PICS used to calculate the 
models. In Figures \ref{fig:g191optcomp1} and \ref{fig:g191optcomp2} we plot the synthetic spectra of models 1 and 2, and 1 
and 3 in the optical region respectively. In both cases, changes to both the continuum flux and H-balmer lines can be seen, 
but these are restricted to $<0.1$\%. The same also occurs for model 4. Because these changes are so small, it is highly 
unlikely that measurements made using these models would be significantly different.

\subsection{Abundance measurements}
In Table \ref{table:newobsabun} we list the abundance measurements made using the various models. We have also given the 
abundance differences between models 1 and 2, 1 and 3, and 1 and 4. Seven ions were found to have statistically significant
abundance differences dependent on which model was used, namely N {\sc v}, O {\sc iv}, O {\sc v}, Fe {\sc iv}, Fe {\sc v}, 
Ni {\sc iv}, and Ni {\sc v}. 

\subsubsection{\normalfont{N {\sc v}}}
For N {\sc v}, significant changes are seen in models 2, 3, and 4. Using model 1, we measured the abundance of N {\sc v} to 
be $1.65^{+0.02}_{-0.02}\times{10}^{-7}$, whereas for models 2, 3, and 4, we find $1.77^{+0.02}_{-0.02}\times{10}^{-7}$, 
$1.87^{+0.02}_{-0.02}\times{10}^{-7}$, and $1.99^{+0.02}_{-0.02}\times{10}^{-7}$ respectively. The abundances measured using 
models 2, 3, and 4 correspond to increases from model 1 of $\sim 7$\%, $\sim 13$\%, and $\sim 20$\% respectively. This suggests that both
the number of transitions and the PICS used cause significant changes to the abundance.

\subsubsection{\normalfont{O {\sc iv}-{\sc v}}}
In the case of O {\sc iv}, significant changes in the abundances are seen when using the {\sc autostructure} 
PICS in models 3 and 4. For O {\sc iv}, we measured the abundance to be $4.63^{+0.12}_{-0.12}\times{10}^{-7}$ using 
model 1. For models 3 and 4, we measured abundances of $4.38^{+0.12}_{-0.12}\times{10}^{-7}$ and 
$4.31^{+0.12}_{-0.12}\times{10}^{-7}$ respectively, corresponding to decreases of $\sim 5$\% and $\sim 7$\% respectively.

A similar case occurs for O {\sc v}, where statistically significant differences occur for models 3 and 4. Using model 1, we 
measured an abundance of $1.47^{+0.07}_{-0.07}\times{10}^{-6}$, whereas for models 3 and 4 we measured abundances of 
$1.69^{+0.08}_{-0.08}\times{10}^{-6}$ and $1.79^{+0.08}_{-0.08}\times{10}^{-6}$ respectively. This corresponds to an increase
of $\sim 15$\% and $\sim 22$\% for models 3 and 4 over model 1 respectively. These results suggest that the largest changes to the 
O {\sc iv}-{\sc v} may be cause by the PICS rather than the number of transitions included in the line list.

\subsubsection{\normalfont{Fe {\sc iv}-{\sc v}}}
The Fe {\sc iv}-{\sc v} abundances appear relatively insensitive to changes in the line list used and the PICS. 
A statistically significant difference was only observed between abundances measured using models 1 and 4. For Fe {\sc iv} we 
measured an abundance of $2.05^{+0.04}_{-0.04}\times{10}^{-6}$ for model 1, and $1.98^{+0.04}_{-0.04}\times{10}^{-6}$ for 
model 4. This is a $\sim 3$\% decrease from model 1. For Fe {\sc v}, we measured abundances of 
$5.37^{+0.08}_{-0.08}\times{10}^{-6}$ and $5.20^{+0.07}_{-0.07}\times{10}^{-6}$ for models 1 and 4 respectively, 
corresponding to a $\sim 3$\% decrease. These changes appear to suggest that a combination of both the number of transitions and 
PICS is required to change the abundance. We discuss this further below.

\subsubsection{\normalfont{Ni {\sc iv}-{\sc v}}}
Interestingly, statistically significant changes compared to model 1 are seen in the Ni {\sc iv} abundance when models 2 and 
3 are used, but not for model 4. For model 1, we measured the abundance to be $3.00^{+0.09}_{-0.04}\times{10}^{-7}$. For 
models 2 and 3, we measure the abundances to be $2.81^{+0.05}_{-0.05}\times{10}^{-7}$ and 
$3.32^{+0.10}_{-0.10}\times{10}^{-7}$ respectively. This corresponds to a decrease of $\sim 6$\% for model 2, and an increase of 
$\sim 11$\% for model 3. In the case of model 4, an abundance of $2.97^{+0.12}_{-0.05}\times{10}^{-7}$ was measured. 

The Ni {\sc v} abundance appears to depend strongly upon whether Ku92 or Ku11 atomic data was included. In model 1, an 
abundance of $9.88^{+0.28}_{-0.21}\times{10}^{-7}$ was measured. For model 2, we measured the abundance to be 
$1.22^{+0.04}_{-0.04}\times{10}^{-6}$, being a $\sim 23$\% increase from model 1. For model 3, only a very small difference was 
noted. We measured the abundance to be $9.81^{+0.23}_{-0.21}\times{10}^{-6}$, which is a $<1$\% decrease. In model 4, we 
again see a large increase in the abundance from model 1, measuring $1.20^{+0.04}_{-0.04}\times{10}^{-6}$, which is a $\sim 21$\% 
increase.

\subsubsection{Ionization fraction agreement}
\begin{figure*}
\begin{centering}
\includegraphics[width=140mm]{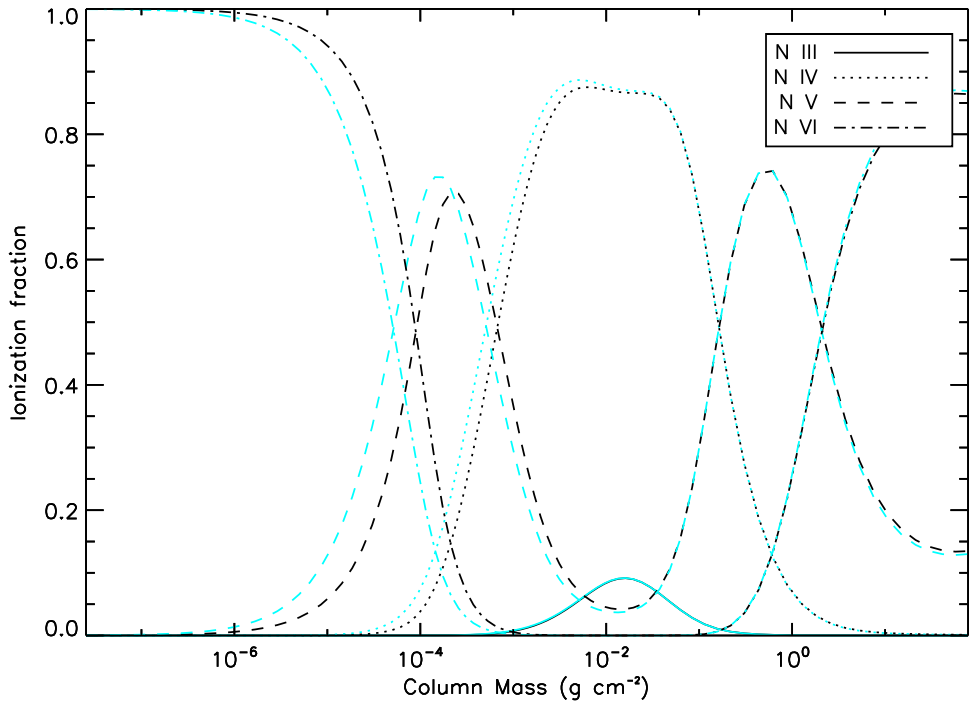}
\caption{Plot of ionization fraction for N {\sc iv}-{\sc vii} for models 1 (black curve) and 3 (cyan curve). Colour figures
are available online.}
\label{fig:ionnitro}
\end{centering}
\end{figure*}

\begin{figure*}
\begin{centering}
\includegraphics[width=140mm]{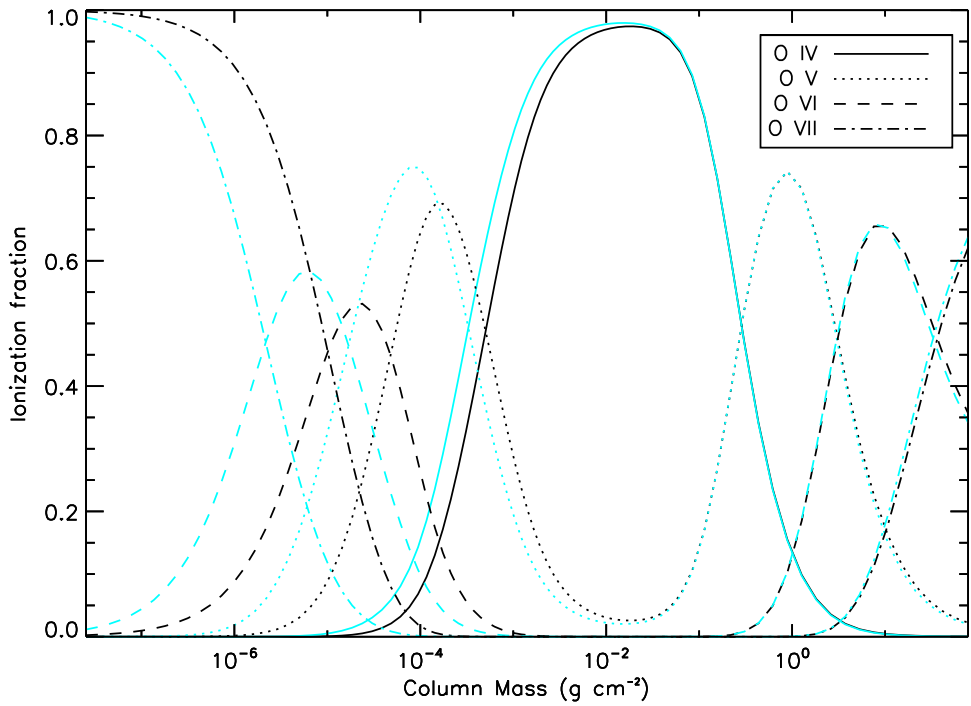}
\caption{Same as Figure \ref{fig:ionnitro}, but for O {\sc iv}-{\sc vii}.}
\label{fig:ionoxy}
\end{centering}
\end{figure*}

In addition to the differences noted above, abundances for different ionization stages of N and O were found to diverge depending 
upon the PICS or atomic data used. In the case of N, the difference between the N {\sc iv} and N {\sc v} abundances for model 1 is $0.02^{+0.22}_{-0.22}\times{10}^{-7}$,
whereas for models 2, 3, and 4 the differences are $-0.12^{+0.22}_{-0.22}\times{10}^{-7}$, $-0.25^{+0.22}_{-0.22}\times{10}^{-7}$, and 
$-0.39^{+0.22}_{-0.22}\times{10}^{-7}$ respectively. For O, the difference between the O {\sc iv} and O {\sc v} abundances for model 1 is $-1.01^{+0.07}_{-0.07}\times{10}^{-6}$,
whereas for models 2, 3, and 4 the differences are $-1.11^{+0.07}_{-0.07}\times{10}^{-6}$, $-1.25^{+0.08}_{-0.08}\times{10}^{-6}$, and 
$-1.36^{+0.08}_{-0.08}\times{10}^{-6}$ respectively. The reason for this can be seen upon inspection of the ionization fractions for N and O.
In Figures \ref{fig:ionnitro} and \ref{fig:ionoxy} we have plotted the ionization fractions for N and O against column mass for models 1 and 3. In both cases it can be seen that
the model 3 ionization fractions have been shifted to smaller column masses. In the case of O, this effect is far more pronounced. In addition, it can also be seen that
the shift to smaller column masses is larger for N/O {\sc v} than it is for N/O {\sc iv}. Therefore, this explains why the abundance measurements diverge. 

Notwithstanding changes to the atomic data and PICS, the overall agreement between abundances measured for different ionization stages of particular
species is generally poor. For this work, we adopted $T_{\mathrm{eff}}=52,500$K and log $g=7.53$ as measured by \cite{barstow03b} for G191-B2B for models 1 to 4.
These values were used for consistency with the work described by \cite{preval13a}. Since then, measurements of $T_{\mathrm{eff}}$ and log $g$ for G191-B2B have been
revised upward by \cite{rauch13a} to 60,000K and 7.60 respectively. The agreement between abundances measured for different ionization stages is a sensitive function
of $T_{\mathrm{eff}}$, atmospheric composition, and to a lesser extent (sans H) log $g$. Our work was not focused on finding the best combination of $T_{\mathrm{eff}}$, log $g$, and atmospheric composition, but instead focused on whether a change, if any, occured to the measured abundances when altering the atomic data and PICS.

\subsubsection{Abundance differences}
Interestingly, the difference between abundances measured using models 1 and 4 can be related to the differences between 
abundances measured using models 1 and 2, and 1 and 3. For example, the difference between the N {\sc v} abundances measured 
using models 1 and 2 is $-0.12^{+0.02}_{-0.02}\times{10}^{-7}$, whereas for models 1 and 3, it is 
$-0.23^{+0.02}_{-0.02}\times{10}^{-7}$. If we add these differences from the abundance found in model 1, we obtain a total of 
$-0.35^{+0.03}_{-0.03}\times{10}^{-7}$. It is for this reason that we include an extra column in Table 
\ref{table:newobsabun}, where the differences between abundances measured using models 1 and 2, and 1 and 3, are summed 
together. It can be seen that in all cases, the sum of these is equal to (within the uncertainties) the 1-4 column. This is 
easily explained in terms of the opacity. Recall that the total opacity in a stellar atmosphere is just a linear sum of each 
individual contribution. In this case, it is the bound-free (PICS) and the bound-bound (Ku92 or Ku11) that is 
being added. Model 2 and Model 3 use the Ku11/Hydrogenic PICS and the Ku92/{\sc autostructure} PICS 
respectively. Given that Model 1 uses the Ku92/Hydrogenic PICS, substracting the abundance found in Model 2 from 
Model 1 shows the effect of including more Ni transitions. Likewise, substracting the abundance found in Model 3 from Model 1 
shows the effect of including more realistic PICS. Therefore, adding these two differences together will give the 
combination of these two effects. This explains why statistically significant differences were observed only when using model 
4 for Fe {\sc iv}-{\sc v}, in that the effects of both the line list and the PICS combine.

\begin{table*}
\renewcommand{\arraystretch}{1.5}
\setlength{\tabcolsep}{3pt}
\centering
\caption{Summary of the abundances measured as a number fraction of H. The 1-2, 1-3, and 1-4 columns give the difference 
between abundances measured using these models. The $\Sigma$ column gives the sum of the 1-2 and 1-3 columns. Differences 
typeset in italics are statistically significant (i.e. consistent with a non-zero difference). Note $[x]=1\times{10}^{x}$}
\begin{tabular}[H]{@{}lcccccccc}
\hline
Ion & Model 1 & Model 2 & Model 3 & Model 4 & 1 - 2 & 1 - 3 & 1 - 4 & $\Sigma$\\
\hline
C {\sc iii}  & $1.83^{+0.03}_{-0.03}[-7]$ & $1.83^{+0.03}_{-0.03}[-7]$ & $1.83^{+0.03}_{-0.03}[-7]$ & $1.83^{+0.03}_{-0.03}[-7]$ & $ 0.00^{+0.04}_{-0.04}[-7]$ & $ 0.00^{+0.04}_{-0.04}[-7]$ & $ 0.00^{+0.04}_{-0.04}[-7]$ & $ 0.00^{+0.06}_{-0.06}[-7]$ \\
C {\sc iv}   & $3.00^{+0.10}_{-0.14}[-7]$ & $3.00^{+0.10}_{-0.17}[-7]$ & $3.00^{+0.10}_{-0.17}[-7]$ & $3.00^{+0.09}_{-0.20}[-7]$ & $ 0.00^{+0.14}_{-0.22}[-7]$ & $ 0.00^{+0.15}_{-0.22}[-7]$ & $ 0.00^{+0.14}_{-0.24}[-7]$ & $ 0.00^{+0.21}_{-0.31}[-7]$ \\
N {\sc iv}   & $1.67^{+0.22}_{-0.22}[-7]$ & $1.65^{+0.22}_{-0.22}[-7]$ & $1.62^{+0.22}_{-0.22}[-7]$ & $1.60^{+0.22}_{-0.22}[-7]$ & $ 0.02^{+0.31}_{-0.31}[-7]$ & $ 0.05^{+0.31}_{-0.31}[-7]$ & $ 0.07^{+0.31}_{-0.31}[-7]$ & $ 0.07^{+0.44}_{-0.44}[-7]$ \\
N {\sc v}    & $1.65^{+0.02}_{-0.02}[-7]$ & $1.77^{+0.02}_{-0.02}[-7]$ & $1.87^{+0.02}_{-0.02}[-7]$ & $1.99^{+0.02}_{-0.02}[-7]$ & $\mathit{-0.12^{+0.02}_{-0.02}[-7]}$ & $\mathit{-0.23^{+0.02}_{-0.02}[-7]}$ & $\mathit{-0.35^{+0.02}_{-0.02}[-7]}$ & $-0.35^{+0.03}_{-0.03}[-7]$ \\
O {\sc iv}   & $4.63^{+0.12}_{-0.12}[-7]$ & $4.54^{+0.12}_{-0.12}[-7]$ & $4.38^{+0.12}_{-0.12}[-7]$ & $4.31^{+0.12}_{-0.12}[-7]$ & $ 0.09^{+0.17}_{-0.17}[-7]$ & $\mathit{ 0.25^{+0.17}_{-0.17}[-7]}$ & $\mathit{ 0.32^{+0.17}_{-0.17}[-7]}$ & $ 0.34^{+0.24}_{-0.24}[-7]$ \\
O {\sc v}    & $1.47^{+0.07}_{-0.07}[-6]$ & $1.56^{+0.07}_{-0.07}[-6]$ & $1.69^{+0.08}_{-0.08}[-6]$ & $1.79^{+0.08}_{-0.08}[-6]$ & $-0.09^{+0.10}_{-0.10}[-6]$ & $\mathit{-0.23^{+0.10}_{-0.10}[-6]}$ & $\mathit{-0.32^{+0.11}_{-0.11}[-6]}$ & $-0.32^{+0.14}_{-0.14}[-6]$ \\
Al {\sc iii} & $1.62^{+0.09}_{-0.09}[-7]$ & $1.63^{+0.10}_{-0.10}[-7]$ & $1.63^{+0.10}_{-0.10}[-7]$ & $1.63^{+0.10}_{-0.10}[-7]$ & $-0.01^{+0.13}_{-0.13}[-7]$ & $-0.01^{+0.13}_{-0.13}[-7]$ & $-0.01^{+0.13}_{-0.13}[-7]$ & $-0.02^{+0.18}_{-0.18}[-7]$ \\
Si {\sc iii} & $2.90^{+0.39}_{-0.23}[-7]$ & $2.89^{+0.38}_{-0.23}[-7]$ & $2.92^{+0.41}_{-0.23}[-7]$ & $2.92^{+0.41}_{-0.23}[-7]$ & $ 0.00^{+0.54}_{-0.32}[-7]$ & $-0.02^{+0.57}_{-0.33}[-7]$ & $-0.02^{+0.56}_{-0.33}[-7]$ & $-0.02^{+0.79}_{-0.46}[-7]$ \\
Si {\sc iv}  & $3.32^{+0.20}_{-0.20}[-7]$ & $3.33^{+0.20}_{-0.20}[-7]$ & $3.34^{+0.20}_{-0.20}[-7]$ & $3.35^{+0.20}_{-0.20}[-7]$ & $-0.01^{+0.28}_{-0.28}[-7]$ & $-0.02^{+0.28}_{-0.28}[-7]$ & $-0.03^{+0.28}_{-0.28}[-7]$ & $-0.03^{+0.40}_{-0.40}[-7]$ \\
P {\sc iv}   & $1.34^{+0.20}_{-0.20}[-7]$ & $1.34^{+0.20}_{-0.20}[-7]$ & $1.30^{+0.19}_{-0.20}[-7]$ & $1.30^{+0.19}_{-0.20}[-7]$ & $ 0.00^{+0.28}_{-0.28}[-7]$ & $ 0.03^{+0.28}_{-0.28}[-7]$ & $ 0.04^{+0.28}_{-0.28}[-7]$ & $ 0.03^{+0.40}_{-0.40}[-7]$ \\
P {\sc v}    & $1.91^{+0.03}_{-0.03}[-8]$ & $1.91^{+0.03}_{-0.03}[-8]$ & $1.91^{+0.03}_{-0.03}[-8]$ & $1.90^{+0.03}_{-0.03}[-8]$ & $ 0.01^{+0.04}_{-0.04}[-8]$ & $ 0.00^{+0.04}_{-0.04}[-8]$ & $ 0.01^{+0.04}_{-0.04}[-8]$ & $ 0.01^{+0.06}_{-0.06}[-8]$ \\
S {\sc iv}   & $2.01^{+0.03}_{-0.03}[-7]$ & $1.99^{+0.03}_{-0.03}[-7]$ & $1.99^{+0.03}_{-0.03}[-7]$ & $1.96^{+0.03}_{-0.03}[-7]$ & $ 0.03^{+0.05}_{-0.05}[-7]$ & $ 0.02^{+0.05}_{-0.05}[-7]$ & $ 0.05^{+0.05}_{-0.05}[-7]$ & $ 0.05^{+0.07}_{-0.07}[-7]$ \\
S {\sc vi}   & $7.55^{+0.20}_{-0.20}[-8]$ & $7.64^{+0.20}_{-0.20}[-8]$ & $7.51^{+0.20}_{-0.20}[-8]$ & $7.59^{+0.20}_{-0.20}[-8]$ & $-0.08^{+0.28}_{-0.28}[-8]$ & $ 0.05^{+0.28}_{-0.28}[-8]$ & $-0.03^{+0.28}_{-0.28}[-8]$ & $-0.03^{+0.40}_{-0.40}[-8]$ \\
Fe {\sc iv}  & $2.05^{+0.04}_{-0.04}[-6]$ & $2.01^{+0.04}_{-0.04}[-6]$ & $2.02^{+0.04}_{-0.04}[-6]$ & $1.98^{+0.04}_{-0.04}[-6]$ & $ 0.04^{+0.05}_{-0.05}[-6]$ & $ 0.03^{+0.05}_{-0.05}[-6]$ & $\mathit{ 0.07^{+0.05}_{-0.05}[-6]}$ & $ 0.07^{+0.07}_{-0.07}[-6]$ \\
Fe {\sc v}   & $5.37^{+0.08}_{-0.08}[-6]$ & $5.31^{+0.07}_{-0.07}[-6]$ & $5.27^{+0.07}_{-0.07}[-6]$ & $5.20^{+0.07}_{-0.07}[-6]$ & $ 0.06^{+0.11}_{-0.11}[-6]$ & $ 0.10^{+0.11}_{-0.11}[-6]$ & $\mathit{ 0.17^{+0.11}_{-0.11}[-6]}$ & $ 0.16^{+0.16}_{-0.16}[-6]$ \\
Ni {\sc iv}  & $3.00^{+0.09}_{-0.04}[-7]$ & $2.81^{+0.05}_{-0.05}[-7]$ & $3.32^{+0.10}_{-0.10}[-7]$ & $2.97^{+0.12}_{-0.05}[-7]$ & $\mathit{ 0.19^{+0.10}_{-0.06}[-7]}$ & $\mathit{-0.32^{+0.13}_{-0.10}[-7]}$ & $ 0.03^{+0.15}_{-0.07}[-7]$ & $-0.13^{+0.16}_{-0.12}[-7]$ \\
Ni {\sc v}   & $9.88^{+0.28}_{-0.21}[-7]$ & $1.22^{+0.04}_{-0.04}[-6]$ & $9.81^{+0.23}_{-0.21}[-7]$ & $1.20^{+0.04}_{-0.04}[-6]$ & $\mathit{-2.30^{+0.29}_{-0.22}[-7]}$ & $ 0.06^{+0.37}_{-0.30}[-7]$ & $\mathit{-2.12^{+0.29}_{-0.22}[-7]}$ & $-2.24^{+0.47}_{-0.37}[-7]$ \\
\hline
\end{tabular}
\label{table:newobsabun}
\renewcommand{\arraystretch}{1.0}
\end{table*}

\subsection{General discussion}
Ideally, any calculation should be as accurate as possible including the most up-to-date data available. However, this also 
needs to be balanced in terms of time constraints, and the task at hand. We have seen that in the EUV, the choice of using 
either Ku92 or Ku11 is irrelevant as the change is very small. The shape of the continuum, however, is very sensitive to the 
PICS used. The downside to using the larger line list from Ku11 increases the calculation time significantly. For 
example, Model 1 took $\sim 17500$ seconds (292 minutes) to converge, whereas Model 2 took $\sim 37000$ seconds (617 minutes). This 
is because the Ku11 data has more energy levels, and is hence split into a larger number of superlevels than for Ku92 data. 

From a wider perspective, the PICS calculated using {\sc autostructure} caused the most changes, in that the EUV 
continuum was severly attenuated, and abundances for N and O were changed. When using the Ku11 line list in model 
atmospheres, abundances for Ni changed significantly while the continua for various spectral regions were left relatively 
unchanged. Given that a calculation with Ku11 data takes twice as long to do than with Ku92, an abundance change in only 
Ni {\sc iv}-{\sc v} is relatively little payoff compared to the physics we can learn from changing the PICS. The 
way forward in improving the quality of future model atmosphere calculations is clear; effort should be focused on improving 
the PICS data for ions where it exists, as well as filling in gaps where it is required (in this case, for Ni).

This piece of work has been a proof-of-concept endeavour. While we have shown that replacing hydrogenic cross-section data 
with more realistic calculations has a significant effect on synthesised spectra and measurements, we have only considered
direct PI. If a direct PI-only calculation has this large an effect, then it stands to reason that a full calculation 
including photoexcitation/autoionization resonances will cause a greater effect. 

The applications of this work is not limited to white dwarf stars. This data can be used in stellar atmosphere models for 
objects of any kind, and any temperature range. We chose to demonstrate the effects of our calculations on a hot DA white 
dwarf star as calculations for these objects are relatively simple. At this temperature regime, we do not need to worry about
the effects of 3D modelling, convection etc.

\subsection{Future work}
As mentioned in our discussion, this work has been a proof-of-concept. The next step is to extend our PICS 
calculations to include other ions of Ni. Once this is done, we plan to include the omitted resonances to our calculations,
and re-examine the effect including this data has on model spectra and measurements. 

The present PICS data was calculated using a distorted wave approximation. A potentially more accurate 
calculation can be achieved using the R-matrix method as in the Opacity Project. Therefore, we aim to do some test 
calculations to compare Ni PICS using both the R-matrix or distorted wave approximation.

The EUV spectra of hot metal polluted white dwarfs has historically been difficult to model (cf. 
\citealt{lanz96a,barstow98a}), the key to which may be the input atomic physics. Therefore, we will also consider the quality 
of fits to the EUV spectra of several metal polluted white dwarf stars.

\section{Conclusion}
We have presented our PICS calculations of Ni {\sc iv}-{\sc vi} using the distorted wave code 
{\sc autostructure}. We investigated the effect of using two different line lists (Ku92 and Ku11) and two different sets of 
PICS (hydrogenic and {\sc autostructure}) on synthesized spectra and abundance measurements based on the hot DA
white dwarf G191-B2B. This investigation was done by calculating four models (labelled 1, 2, 3, and 4) with permutations of 
the Ni line list and PICS used. Model 1 used Ku92 line list/hydrogenic PICS, Model 2 used Ku11 line 
list/hydrogenic PICS, Model 1 used Ku92 line list/{\sc autostructure} PICS, and model 4 used Ku11 line 
list/{\sc autostructure} PICS.

We synthesised model spectra for each of the four models in the EUV, UV, and optical regions. In the EUV, model 3 showed
large attentuation shortward of 180\AA\, of up to $\sim 80$\% relative to model 1, whereas model 2 was relatively unchanged. 
In the UV, the continuum was unchanged in models 2 and 3. However, in model 2, the Ni absorption feature depths changed 
significant, increasing in depth by up to $\sim 10$\%. Absorption features in model 3 were relatively unchanged, with depth 
changes of $\sim 3$\% across the spectrum. In the optical, changes in flux were so small ($<0.1$\%) across models that these 
are unlikely to be observed, nor would it be possible to differentiate between them. Model 4 was not plotted in the EUV, UV, 
or optical as the resultant spectrum was just a combination of the effects observed in models 2 and 3.

We measured metal abundances for G191-B2B using all four models. This was to see if there were any differences in the 
metal abundances measured when changing the PICS or atomic data included in the model calculation. Statistically significant (consistent with non-zero 
difference compared to model 1) abundance changes were observed in N {\sc v} over all models, and O {\sc iv}-{\sc v} when 
using models 3 and 4. This suggests that the N abundances are sensitive to both the line list and PICS used, while 
the O abundances were only sensitive to the PICS. The Fe {\sc iv}-{\sc v} abundances only changed by a 
statistically significant amount for model 4, implying a combination of the line list and cross-section caused the change. 
The Ni {\sc iv}-{\sc v} abundances changed by a statistically significant amount for models 2 and 4, implying the line list
caused the difference. Interestingly, for each metal abundance, the difference between measurements made using models 1 and 4
could be found by summing the differences between measurements made using models 1 and 2, and models 1 and 3. This is in 
keeping with the assumption that predicted radiation for small variations of the opacity sources scales roughly linearly with 
the opacity. 

In addition, we found the abundances of N/O {\sc iv} and {\sc v} diverged depending on the PICS used. A comparison of the 
ionization fractions calculated using models 1 and 3 showed that the charge states for model 3 formed higher in the
atmosphere than the charge states for model 1. Furthermore, N/O {\sc v} experiences a larger change with respect to 
depth formation than N/O {\sc iv}, explaining the divergence of the abundance measurements.

Our work has demonstrated that, even with a limited calculation, the Ni PICS have made a significant difference
to the synthetic spectra, and by extension what is measured from observational data. Comparatively, an extended line list 
such as Ku11 offers little benefit given the extended computational time required to converge a model atmosphere, and the 
small pay off (changed Ni abundance). It is our opinion that future atomic data calculations for stellar atmosphere models 
should not necessarily focus on how big the line list is, but the quality of the PICS.

\section*{Acknowledgments}
We gratefully acknowledge the time and effort expended by Robert Kurucz in assisting with this project. We also thank Simon 
Jeffery and Nigel Bannister for helpful discussions. SPP and MAB acknowledge the support of an STFC student grant. 



\bibliographystyle{mn2e}
\bibliography{simonpreval}

\begin{thebibliography}{}

\bibitem[\protect\citeauthoryear{{Anderson}}{{Anderson}}{1989}]{anderson89a}
{Anderson} L.~S.,  1989, ApJ, 339, 558

\bibitem[\protect\citeauthoryear{{Arnaud}}{{Arnaud}}{1996}]{arnaud96a}
{Arnaud} K.~A.,  1996, in {Jacoby} G.~H.,  {Barnes} J.,  eds, Astronomical Data
  Analysis Software and Systems V Vol.~101 of Astronomical Society of the
  Pacific Conference Series, {XSPEC: The First Ten Years}.
p.~17

\bibitem[\protect\citeauthoryear{Badnell}{Badnell}{1986}]{badnell86a}
Badnell N.~R.,  1986, J. Phys. B, 19, 3827

\bibitem[\protect\citeauthoryear{Badnell}{Badnell}{1997}]{badnell97a}
Badnell N.~R.,  1997, J. Phys. B, 30, 1

\bibitem[\protect\citeauthoryear{Badnell}{Badnell}{2011}]{badnell11a}
Badnell N.~R.,  2011, Comput. Phys. Commun., 182, 1528

\bibitem[\protect\citeauthoryear{Barstow, Good, Holberg, Hubeny, Bannister,
  Bruhweiler, Burleigh \& Napiwotzki}{Barstow et~al.}{2003}]{barstow03b}
Barstow M.~A.,  Good S.~A.,  Holberg J.~B.,  Hubeny I.,  Bannister N.~P.,
  Bruhweiler F.~C.,  Burleigh M.~R.,    Napiwotzki R.,  2003, MNRAS, 341, 870

\bibitem[\protect\citeauthoryear{Barstow, Hubeny \& Holberg}{Barstow
  et~al.}{1998}]{barstow98a}
Barstow M.~A.,  Hubeny I.,    Holberg J.~B.,  1998, MNRAS, 299, 520

\bibitem[\protect\citeauthoryear{{Chayer}, {Fontaine} \& {Wesemael}}{{Chayer}
  et~al.}{1995}]{chayer95a}
{Chayer} P.,  {Fontaine} G.,    {Wesemael} F.,  1995, ApJS, 99, 189

\bibitem[\protect\citeauthoryear{Cowan}{Cowan}{1981}]{cowanbook1981}
Cowan R.~D.,  1981, {The Theory of Atomic Structure and Spectra}.
Los Alamos Series in Basic and Applied Sciences, University of California Press

\bibitem[\protect\citeauthoryear{Eissner \& Nussbaumer}{Eissner \&
  Nussbaumer}{1969}]{eissner69a}
Eissner W.,  Nussbaumer H.,  1969, J. Phys. B, 2, 1028

\bibitem[\protect\citeauthoryear{Holberg, Hubeny, Barstow, Lanz, Sion \&
  Tweedy}{Holberg et~al.}{1994}]{holberg94a}
Holberg J.~B.,  Hubeny I.,  Barstow M.~A.,  Lanz T.,  Sion E.~M.,    Tweedy
  R.~W.,  1994, ApJL, 425, L105

\bibitem[\protect\citeauthoryear{Hubeny}{Hubeny}{1988}]{hubeny88a}
Hubeny I.,  1988, Comput. Phys. Commun., 52, 103

\bibitem[\protect\citeauthoryear{Hubeny \& Lanz}{Hubeny \&
  Lanz}{1995}]{hubeny95a}
Hubeny I.,  Lanz T.,  1995, ApJ, 439, 875

\bibitem[\protect\citeauthoryear{Hubeny \& Lanz}{Hubeny \&
  Lanz}{2011}]{hubeny11a}
Hubeny I.,  Lanz T.,  2011, in Astrophysics Source Code Library, record
  ascl:1109.022 {Synspec: General Spectrum Synthesis Program}.
p.~9022

\bibitem[\protect\citeauthoryear{{Kurucz}}{{Kurucz}}{1992}]{kurucz92a}
{Kurucz} R.~L.,  1992, RMxAA, 23, 45

\bibitem[\protect\citeauthoryear{Kurucz}{Kurucz}{2011}]{kurucz11a}
Kurucz R.~L.,  2011, Canadian Journal of Physics, 89, 417

\bibitem[\protect\citeauthoryear{{Lanz}, {Barstow}, {Hubeny} \&
  {Holberg}}{{Lanz} et~al.}{1996}]{lanz96a}
{Lanz} T.,  {Barstow} M.~A.,  {Hubeny} I.,    {Holberg} J.~B.,  1996, ApJ, 473,
  1089

\bibitem[\protect\citeauthoryear{{Preval}, {Barstow}, {Holberg} \&
  {Dickinson}}{{Preval} et~al.}{2013}]{preval13a}
{Preval} S.~P.,  {Barstow} M.~A.,  {Holberg} J.~B.,    {Dickinson} N.~J.,
  2013, MNRAS, 436, 659

\bibitem[\protect\citeauthoryear{{Rauch}, {Werner}, {Bohlin} \& {Kruk}}{{Rauch}
  et~al.}{2013}]{rauch13a}
{Rauch} T.,  {Werner} K.,  {Bohlin} R.,    {Kruk} J.~W.,  2013, A\&A, 560

\bibitem[\protect\citeauthoryear{{Seaton} \& {Badnell}}{{Seaton} \&
  {Badnell}}{2004}]{seaton2004a}
{Seaton} M.~J.,  {Badnell} N.~R.,  2004, MNRAS, 354, 457

\bibitem[\protect\citeauthoryear{Vennes \& Lanz}{Vennes \&
  Lanz}{2001}]{vennes01a}
Vennes S.,  Lanz T.,  2001, ApJ, 553, 399

\bibitem[\protect\citeauthoryear{Werner \& Dreizler}{Werner \&
  Dreizler}{1994}]{werner94a}
Werner K.,  Dreizler S.,  1994, AAP, 286, L31

\end{thebibliography}

\bsp	
\label{lastpage}
\end{document}